\documentclass[a4paper]{jpconf}
\usepackage{graphicx}
\usepackage{booktabs}
\begin{document}
\title{Ensemble spectral variability study of Active Galactic Nuclei from the XMM-Newton serendipitous source catalogue}

\author{R. Serafinelli, F. Vagnetti, R. Middei}

\address{Dipartimento di Fisica, Universit\`a di Roma ``Tor Vergata'', Via della Ricerca
Scientifica 1, 00133 Roma, Italy}

\ead{roberto.serafinelli@roma2.infn.it}

\begin{abstract}
The variability of the X-ray spectra of active galactic nuclei (AGN) usually includes a change of the spectral slope. This has been investigated for a small sample of  local AGNs by Sobolewska and Papadakis, who found that slope variations are well correlated with flux variations, and that spectra are typically steeper in the bright phase (\textit{softer when brighter} behaviour). Not much information is available for the spectral variability of high-luminosity AGNs and quasars. In order to investigate this phenomenon, we use data from the XMM-Newton Serendipitous Source Catalogue, Data Release 5, which contains X-ray observations for a large number of active galactic nuclei in a wide luminosity and redshift range, for several different epochs. This allows to perform an ensemble analysis of the spectral variability for a large sample of quasars. We quantify the spectral variability through the \textit{spectral variability parameter} $\beta$, defined as the ratio between the change in spectral slope and the corresponding logarithmic flux variation. We find that the spectral variability of quasars has a \textit{softer when brighter} behaviour, similarly to local AGNs.

\end{abstract}

\section{Introduction}
Active galactic nuclei are very bright extragalactic objects, located in the dynamical center of their host galaxy, powered by the accretion of matter around a supermassive ($M_{BH}=10^6$-$10^9 M_{\odot}$) black hole, along a geometrically thin, optically thick disk. This disk is believed to be responsible for the optical/UV emission, while a hot ($T\simeq10^8$-$10^9$ K) corona is believed to produce hard X-ray photons by comptonization of the disk photons. This emission is responsible for the observed power-law like spectrum in the X-ray band.\\
Variability can be a powerful instrument to probe this physical effect, because it can provide information about both shape and physical behaviour of the corona, which are far from being completely understood.\\
In addition to the study of amplitude variability, it is also interesting to investigate the spectral variability of active galactic nuclei. This feature has already been studied in the optical/UV band by many authors, (e.g.  [$1$-$3$]) and a \textit{harder when brighter} behaviour of a power-law spectrum $F\propto \nu^{\alpha}$ has been observed. We use the \textit{spectral variability parameter} $\beta$ introduced in \cite{trevese02}, defined as:

\begin{equation}
\label{eq:betaalpha}
\beta=\frac{\Delta\alpha}{\Delta\log{F}}\, ,
\end{equation}

\noindent
relating variations in flux and spectral index $\alpha$.

\noindent
In the X-ray band the opposite behaviour (\textit{softer when brighter}) has been found for some individual sources (e.g. \cite{magdziarz98, zdiarski03} and more recently \cite{ursini15}). One of the few systematic studies of X-ray spectral variability has been performed by Sobolewska and Papadakis \cite{sobolewska09}, who found such behaviour for $10$ nearby Seyfert galaxies, hinting that most of the sources can match the scenario of the superposition of a constant reflection component that overlaps a power-law component, variable in both amplitude and slope.\\
Currently there have been no systematic studies concerning the spectral variability of quasars and high-luminosity type-$1$ AGNs, one notable exception being \cite{paolillo04}, in which the authors find evidence of spectral variability for a fraction of the sources, also with a softer when brighter trend.\\
The paper is organized as follows: Section $2$ describes the data extracted from the archival catalogues, Section $3$ describes the relations between the beta parameter and the hardness ratio variations, Section $4$ discusses the results of our analysis.

\section{The Dataset}
\label{sec:dataset}
We are looking for the \textit{softer when brighter} effect also in type-$1$ luminous AGNs and quasars.\\
In order to select this kind of sources, we have cross-matched the XMM-Newton Serendipitous Source Catalogue (XMMSSC-DR5 \cite{rosen15}) with two quasar catalogues from the Sloan Digital Sky Survey, i.e. SDSS-DR7Q \cite{schneider10} and SDSS-DR12Q \cite{paris15}. These two optical catalogues are complementary and therefore both must be included, so that the catalogue we are going to analyze could be as complete as possible. SDSS-DR7Q lists $105,783$ spectroscopically confirmed quasars and SDSS-DR12Q lists $297,301$ quasars. Both catalogs do not include active galactic nuclei classes such as Seyfert galaxies, BL Lacertae and type-2 quasars. Therefore, they are well suited for our goal.\\
XMMSSC-DR5 contains $565,962$ observations taken between $2000$ and $2013$, corresponding to $396,910$ unique sources, which means that many sources have been observed more than once. Since we are dealing with variability issues, we required that the sources must be observed multiple times, at least twice. Therefore we only selected sources with multi-epoch observations.\\

\begin{table}[h]
  \begin{center}
    \begin{tabular}{ccc}
      \toprule
      Catalogue & Number of observations & Number of sources\\
      \midrule
      SDSS-DR7Q & / & $105,783$\\
      SDSS-DR12Q & / & $297,301$\\
      XMMSSC-DR5 & $565,962$ & $396,910$\\
      Cross-matched catalogue & $7,837$ & $2,700$\\
      \bottomrule
    \end{tabular}
    \caption{\label{tab:tabnobsnsour}Number of observations and sources listed for all the catalogues used in this work.\\}
  \end{center}
\end{table}

\noindent
Both SDSS catalogues and XMM-Newton Serendipitous Source Catalogue list equatorial coordinates $\alpha$ and $\delta$ for each observation. We included in our catalogue every XMMSSC observation that lies within $5"$ from any catalogued source in any of the two SDSS quasar catalogues. In the cases in which we found the same source in both SDSS catalogues, we chose the one contained in SDSS-DR12Q, because it is more recent. \\
Something that we had to take into account is that not all of the XMMSSC photometric observations are optimal: the quality of the observation is quantified by a parameter of the catalogue, labeled as  \verb|SUM_FLAG|. We excluded from our list the worst photometric observations, meaning all observations with \verb|SUM_FLAG| $>2$ and we obtained a final catalogue that has $7,837$ X-ray observations, corresponding to $2,700$ quasar sources.\\
The number of observations and the number of sources listed in each catalogue are summarized in Table \ref{tab:tabnobsnsour}. \\

\section{The $\beta$ Parameter and the Hardness Ratios}
\label{sec:betapar}
We quantify the spectral variability of a source by means of the \textit{spectral variability parameter} $\beta$ \cite{trevese02}, defined by Eq. \ref{eq:betaalpha}.\\
In the X-ray band, though, the \textit{photon index} $\Gamma$ is more commonly adopted, with the power-law spectra described as $n(E)\propto E^{-\Gamma}$. $\Gamma$ is thus related to $\alpha$ by $\Gamma=-\alpha+1$; this means that the spectral variability parameter is:

\begin{equation}
\label{eq:beta}
\beta=-\frac{\Delta\Gamma}{\Delta\log{F}}\, .
\end{equation}

\noindent
Unfortunately, the photon index $\Gamma$ is not available. However, it is related to the \textit{hardness ratio}, and thus it is useful to look for correlations between the hardness ratios and the fluxes of the ensemble set.

\subsection{HR-Flux Correlations}
Among the many quantities presented, the XMMSSC-DR5 catalogue lists the integrated flux at five different bands, labeled from \verb|EP1_FLUX| to \verb|EP5_FLUX| (that we rename as $F_{1}$ to $F_{5}$ for simplicity), respectively in the bands $0.2$-$0.5$ keV, $0.5$-$1$ keV, $1$-$2$keV, $2$-$4.5$ keV and $4.5$-$12$ keV. XMMSSC-DR5 also lists \textit{hardness ratios} between fluxes in different bands, defined as follows:

\begin{equation}
\label{eq:hardnessratio}
HR_{i}=\frac{CR_{i+1}-CR_{i}}{CR_{i+1}+CR_{i}}\, ,
\end{equation}

\noindent
where $CR_{i}$ is the countrate in the band $i$ and is related to the flux $F_{i}$ by the relation 

\begin{equation}
\label{eq:fcr}
F_{i}=a_{i}CR_{i}\, ,
\end{equation}
\noindent
with the conversion factors $a_i$ provided by the on-line utility \verb|WebPIMMS|\footnote{https://heasarc.gsfc.nasa.gov/cgi-bin/Tools/w3pimms/w3pimms.pl}.

\begin{figure}[ht]
\begin{center}
\includegraphics[height=40mm, width=50mm]{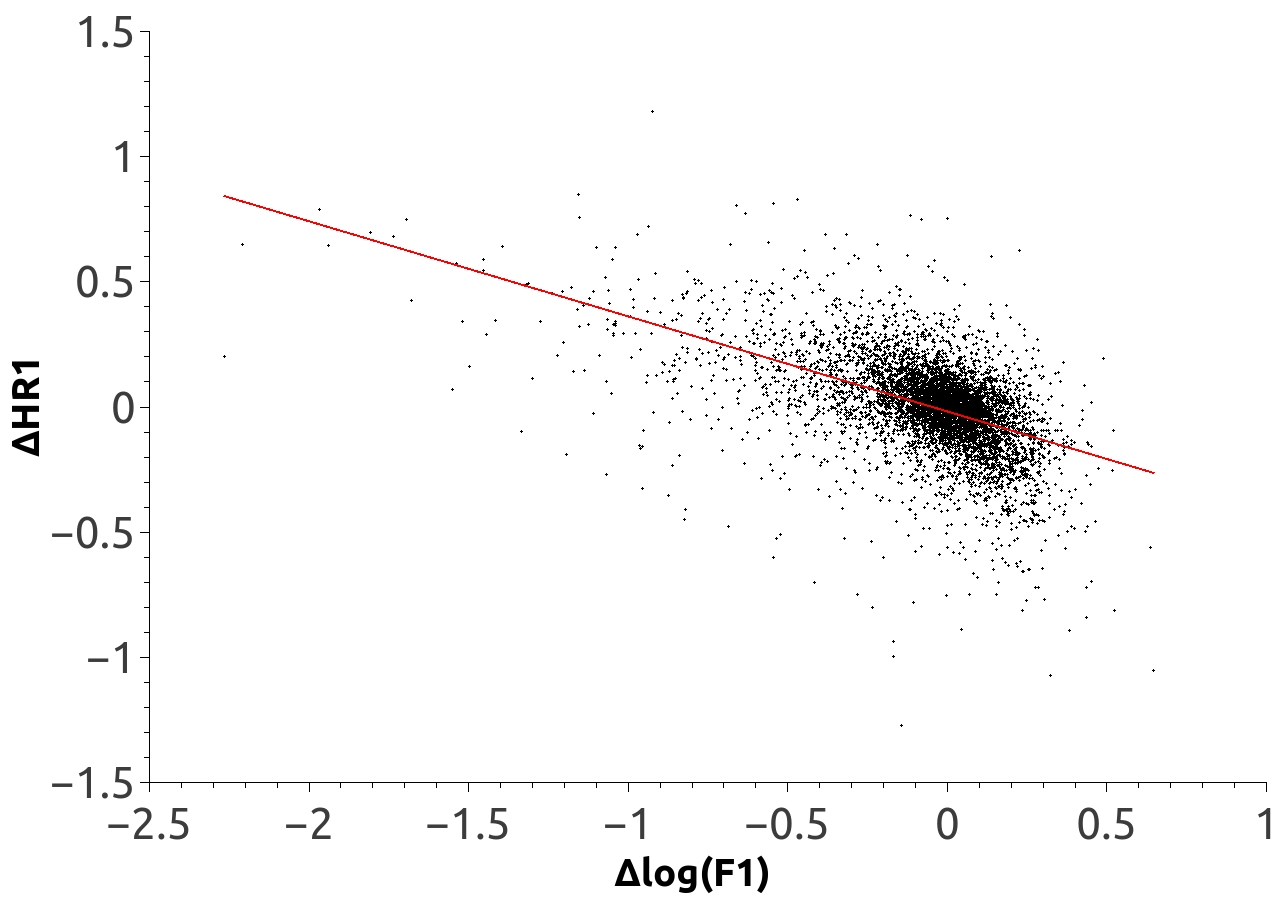}
\includegraphics[height=40mm, width=50mm]{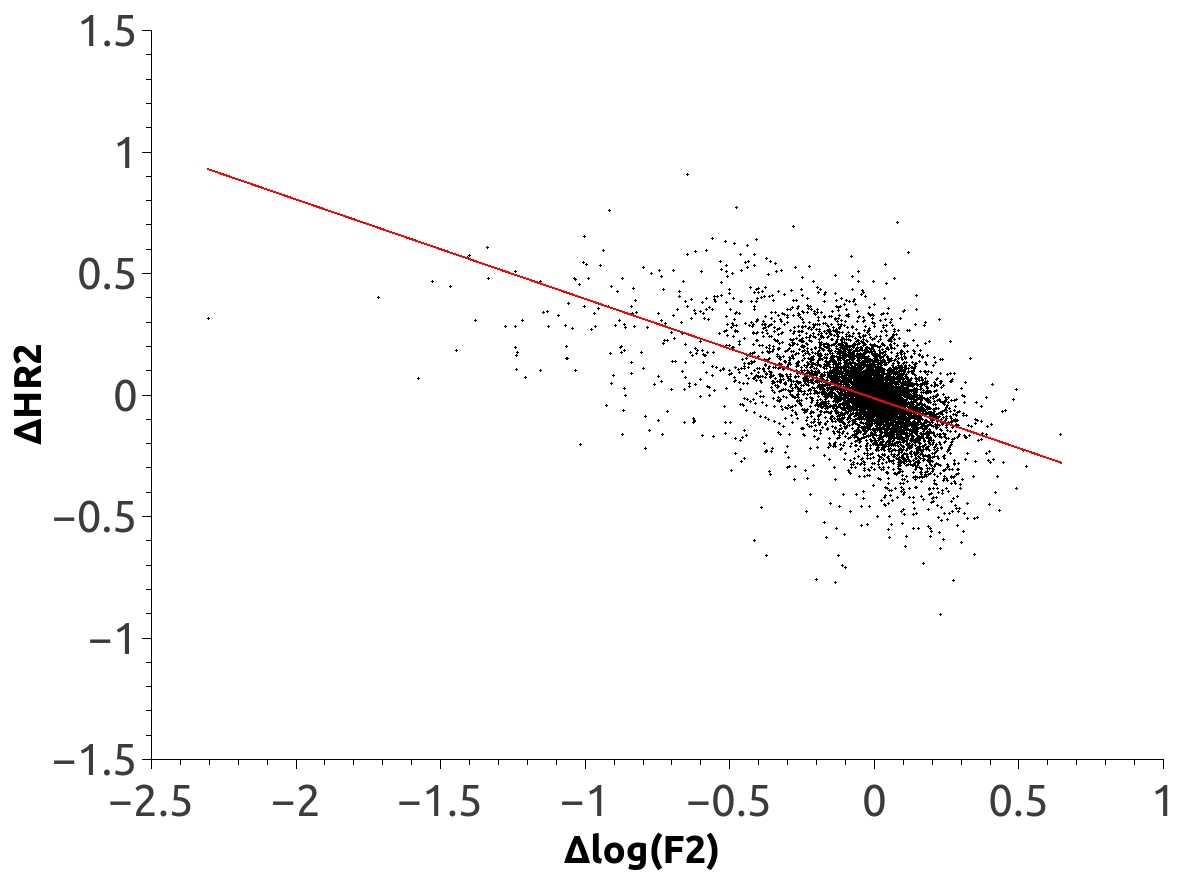}
\includegraphics[height=40mm, width=50mm]{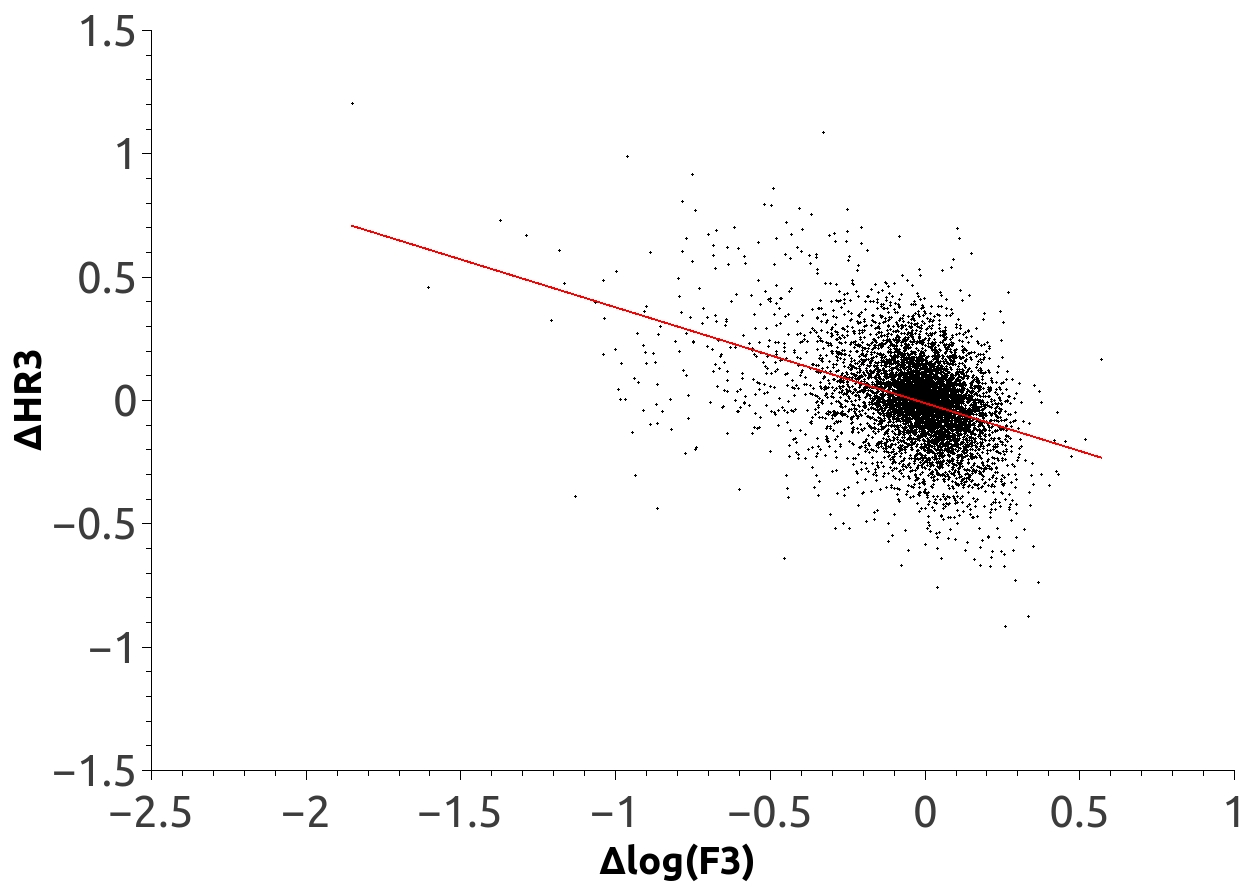}
\caption{\label{fig:fitsgraph} Variations of hardness ratio plotted versus the variations of $\log{F}$ for the three considered bands. The red line is the least-squares best fit for a linear trend.}
\end{center}
\end{figure}

\noindent
\\

\begin{table}
\begin{center}
\begin{tabular}{ccc}
\toprule
Band considered & $\frac{\Delta HR_{i}}{\Delta\log{F_{i}}}$ & Correlation coefficient $r$ \\
\midrule
$\Delta HR_{1}$ vs $\Delta$log$F_{1}$ & $-0.380\pm 0.008$ & $-0.508$\\
$\Delta HR_{2}$ vs $\Delta$log$F_{2}$ & $-0.409\pm 0.008$ & $-0.507$\\
$\Delta HR_{3}$ vs $\Delta$log$F_{3}$ & $-0.388\pm 0.011$ & $-0.384$\\
\bottomrule
\end{tabular}
\caption{\label{tab:fits} Slopes and correlation coefficients obtained from $\Delta HR$ vs $\Delta$log$F$ linear fits.}
\end{center}
\end{table}

\noindent
We also had to take into account that different sources have different average fluxes and hardness ratios among each other. Thus, instead of correlations between $HR_{i}$ and log$F_{i}$, we studied the variations from the mean value of each source of hardness ratio, and the variations, also from the mean value of each source, of the flux logarithm.\\
We have calculated linear fits for $\Delta HR_{1,2,3}$ vs $\Delta$log$F_{1,2,3}$, respectively. Table \ref{tab:fits} shows the slopes obtained for each linear fit and the correlation coefficient, while Fig. \ref{fig:fitsgraph} shows the three fits in the $\Delta HR_i$-$\Delta$log$F_i$ planes. The probabilities of finding these correlations by chance are negligible, given the high number of points involved.

\subsection{$\beta$-HR relation}
As we hinted earlier in this section, the spectral variability parameter $\beta$ can be related to the hardness ratio:

$$\frac{\Delta HR}{\Delta \log{F}}\simeq \frac{d HR}{d \Gamma}\frac{\Delta\Gamma}{\Delta\log{F}}.$$
Using Eq. \ref{eq:beta} we obtain that
$$\frac{\Delta HR}{\Delta \log{F}}\simeq -\beta\frac{d HR}{d \Gamma},$$
which means

\begin{equation}
\label{eq:betagamma}
\beta\simeq -\frac{\frac{\Delta HR}{\Delta \log{F}}}{\frac{dHR}{d\Gamma}}\, .
\end{equation}

\section{Data Analysis and Results}
In order to estimate the spectral variability parameter $\beta$ we need to consider Eq. \ref{eq:betagamma}.\\
The numerator can be taken from Table \ref{tab:fits}. It should be stressed that this is an average ensemble estimate, and that the spectral variability behaviour can be different from source to source.\\
Given the absence of the photon index $\Gamma$ from our catalogue, we proceeded with simulated data in order to numerically compute the denominator of Eq. \ref{eq:betagamma} as a function of $\Gamma$. We have made use of the \verb|X-SPEC| v.12.9.0 software package [12] to generate some spectral models that could fit our typical sources, using the command \verb|model zwabs*powerlaw|, that also takes redshift into account. According to the standard unified model of AGN [13], type-$1$ AGNs are unobscured, since they are face-on oriented, which means that the obscuring torus does not affect this kind of active galaxies. One of the spectral models we have considered, therefore, is the unobscured power-law model. Although this is the most likely scenario, one cannot exclude that X-ray obscuration processes are at work, indipendently from optical/UV obscuration processes. Thus, we have also studied some models that include a moderate absorption over a power-law spectrum. We have only considered models with $\log{N_H}$=21 and $\log{N_H}=22$. We generated models with values of $\Gamma$ ranging from $1$ to $3$, the typical values for AGNs, and redshift $z$ between $0$ and $3$.
For each model we have computed the integrated fluxes in bands $0.2$-$0.5$ keV, $0.5$-$1$ keV, $1$-$2$ keV and $2$-$4.5$ keV (or $F_1$ to $F_4$ for short), necessary to compute the simulated hardness ratios $HR_{1}$, $HR_{2}$ and $HR_{3}$.\\
In fact, using Eq. \ref{eq:hardnessratio} and Eq. \ref{eq:fcr}, and defining $k_{i}=1/a_{i},$ we can write the hardness ratio as:

\begin{equation}
\label{eq:hrflux}
HR_{i}=\frac{k_{i+1} F_{i+1}-k_{i}F_{i}}{k_{i+1} F_{i+1}+k_{i}F_{i}}\, .
\end{equation}

\begin{figure}[ht]
\begin{center}
\includegraphics[height=40mm, width=50mm]{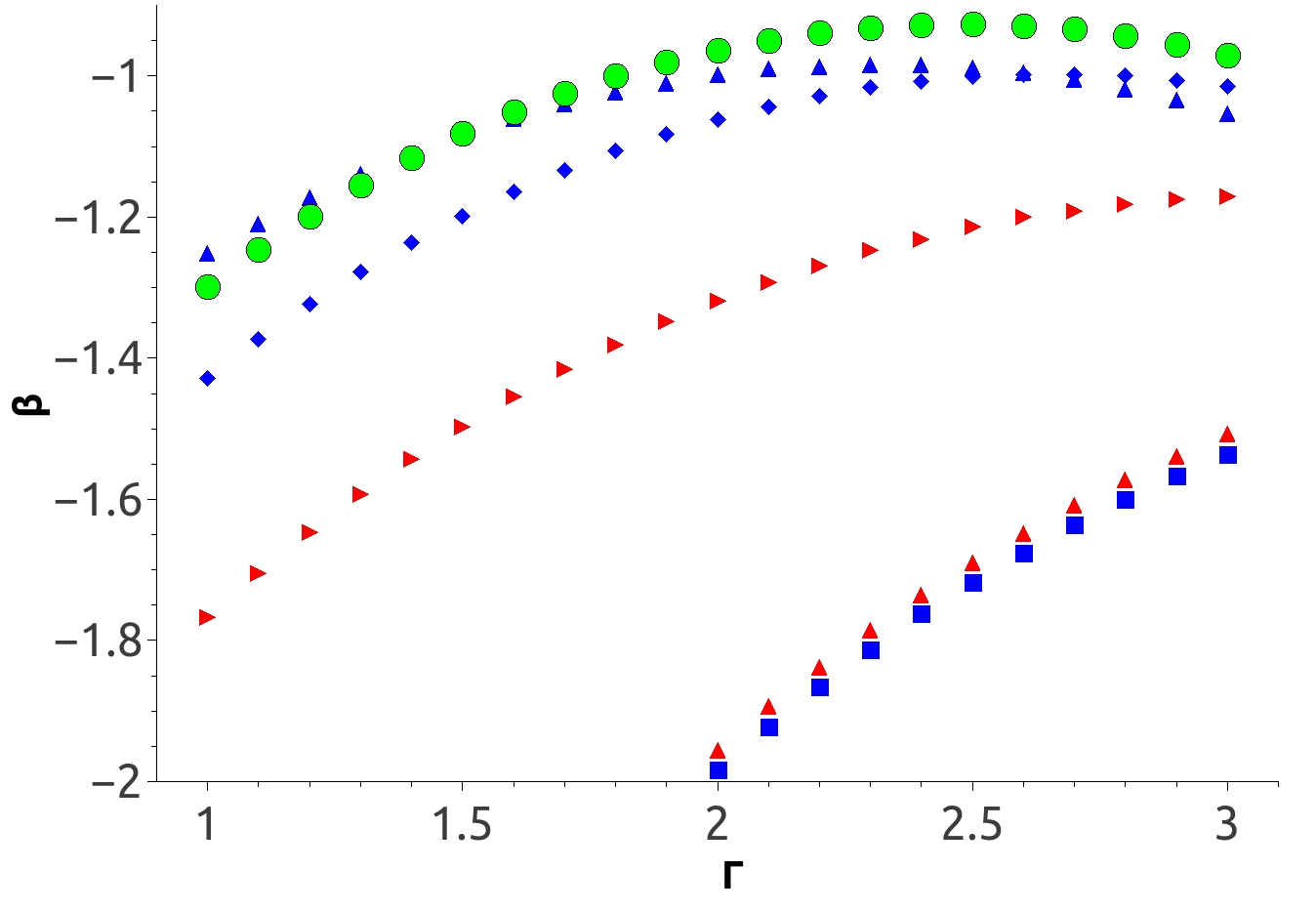}
\includegraphics[height=40mm, width=50mm]{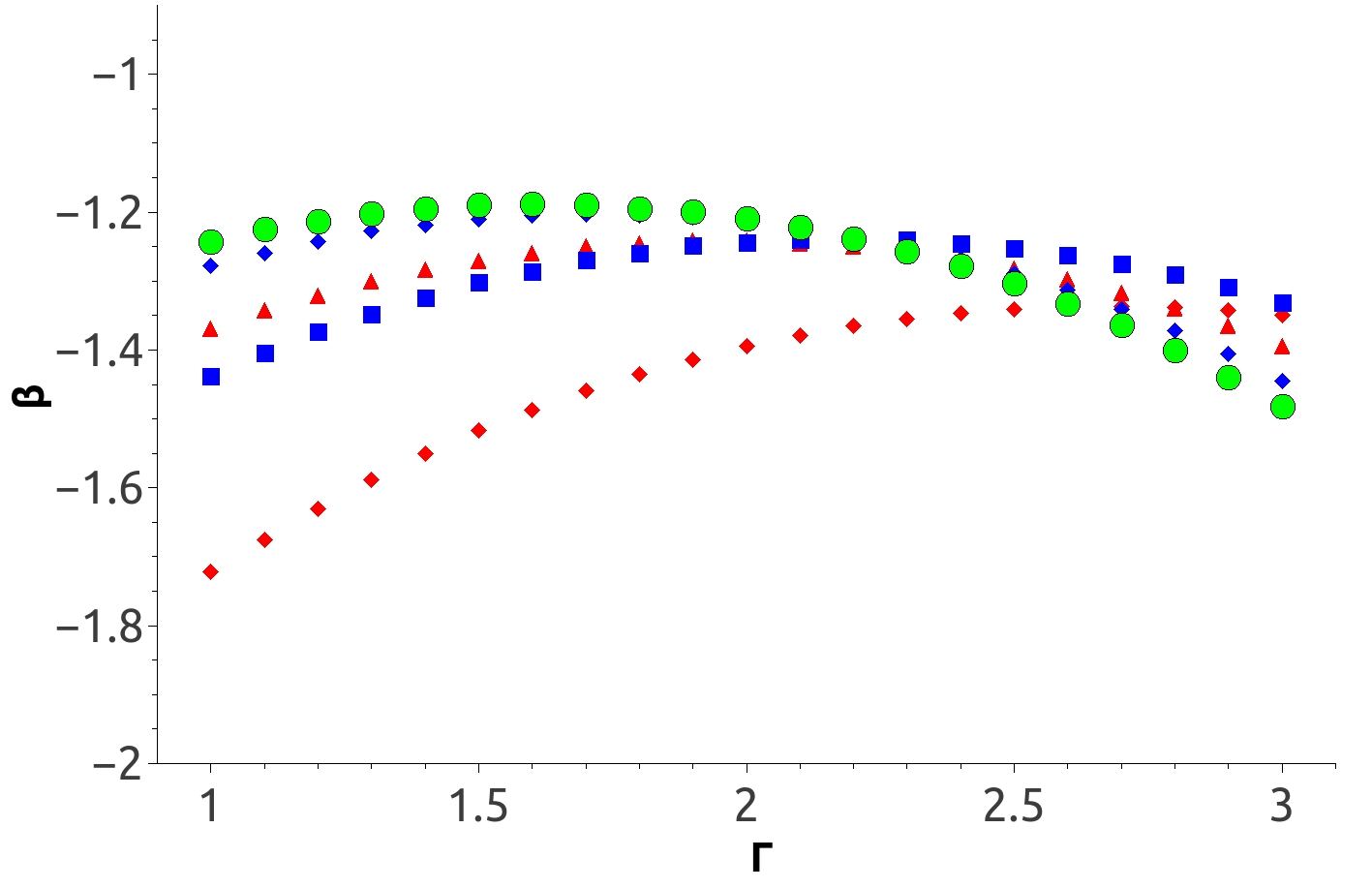}
\includegraphics[height=40mm, width=50mm]{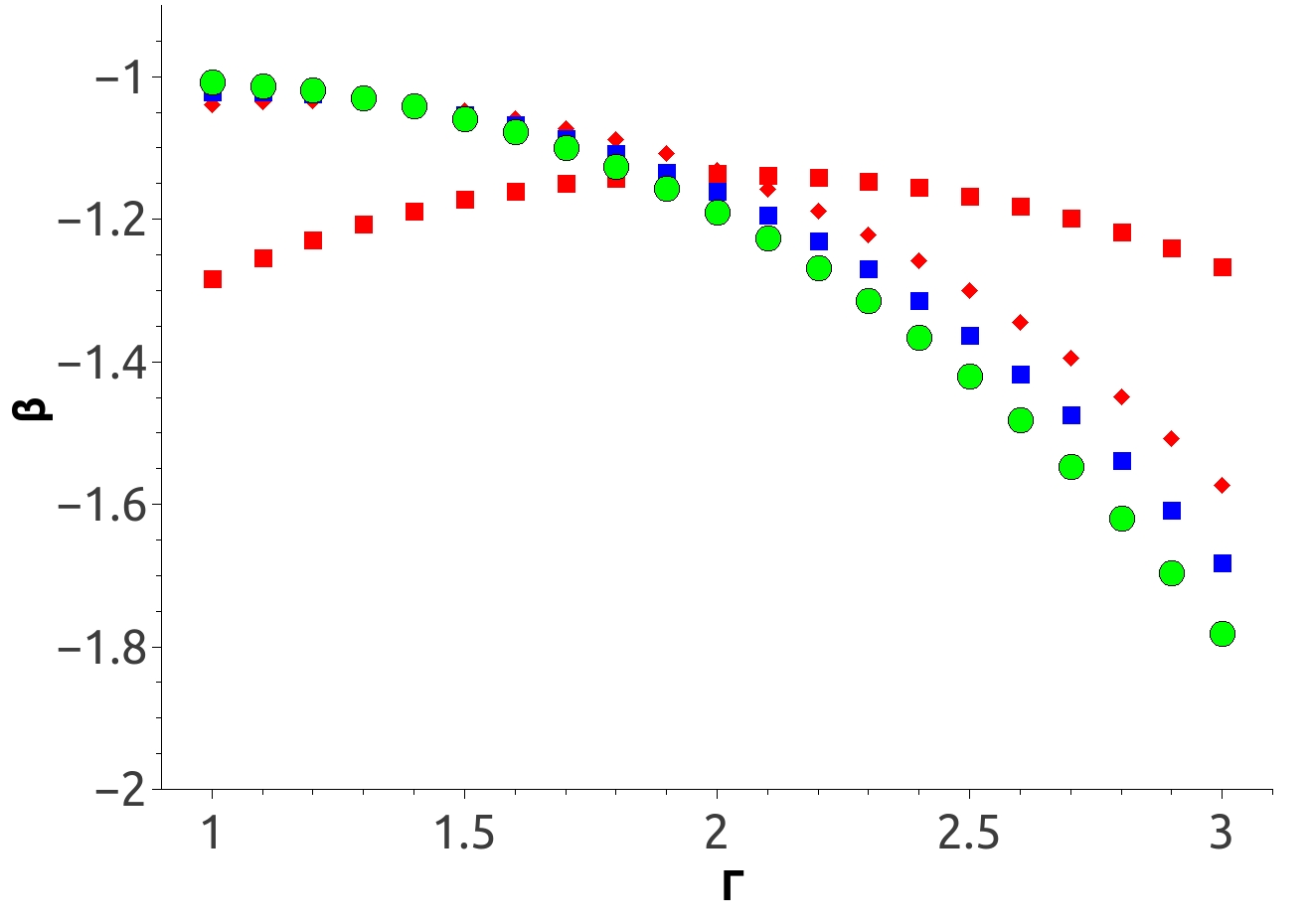}
\caption{\label{fig:models} $\beta$ vs $\Gamma$ plots for some of the models studied, computed using $HR_1$-$F_1$, $HR_2$-$F_2$ and $HR_3$-$F_3$ correlations, from left to right. The green circles represent the unabsorbed power-law model, while moderate absorbed power-law models are represented by blue ($\log{N_H}=21$) and red ($\log{N_H}=22$) symbols. Squares, diamonds, up-pointing triangles and right-pointing triangles refer to redshift $z=0,1,2,3$ respectively.}
\end{center}
\end{figure}

\noindent
We were able to compute therefore, for each model, the values of $HR_{1,2,3}$ and the corresponding values of $\beta$, by computing the numerical derivatives of $HR$ over $\Gamma$.\\
For typical values of $\Gamma$ in the range $1$-$3$, this parameter has negative values (see Fig. \ref{fig:models}), which means that a \textit{softer when brighter} behaviour is globally present. It is to be stressed that these results extend the trend previously discovered for local sources to a wide redshift and luminosity range.

\section*{Acknowledgements}
We thank the organizers of the sixth Young Researchers Meeting. We thank the Gran Sasso Science Institute for their hospitality. This research has made use of data obtained from the 3XMM XMM-Newton serendipitous source catalogue compiled by the $10$ institutes of the XMM-Newton Survey Science Centre selected by ESA.


\section*{References}
\begin{thebibliography}{9}
\bibitem{trevese02} D. Trevese and F. Vagnetti, 2002, \textit{ApJ}, \textbf{564}, 624.
\bibitem{giveon99} U. Giveon, D. Maoz, S. Kaspi et al., 1999, \textit{MNRAS}, \textbf{306}, 637.
\bibitem{vandenberk03} D. E. Vanden Berk, B. C. Wilhite, R. G. Kron et al., 2003, \textit{ApJ}, \textbf{601}, 692.
\bibitem{magdziarz98} P. Magdziarz, O. M. Blaes, A. A. Zdziarski et al., 1998, \textit{MNRAS}, \textbf{301}, 179.
\bibitem{zdiarski03} A. A. Zdziarski, P. Lubinski, M. Gilfanov, M. Revnivtsev, 2003, \textit{MNRAS}, \textbf{342}, 355.
\bibitem{ursini15} F. Ursini, P. O. Petrucci, G. Matt et al., 2015, \texttt{arXiv:1510.01966}.
\bibitem{sobolewska09} M. A. Sobolewska and I. E. Papadakis, 2009, \textit{MNRAS}, \textbf{399}, 1597.
\bibitem{paolillo04} M. Paolillo, E. J. Schreier, R. Giacconi et al., 2004, \textit{ApJ}, \textbf{611}, 93.
\bibitem{rosen15} S. R. Rosen, N. A. Webb, M. G. Watson et al., 2015, \texttt{arXiv:1504.07051}.
\bibitem{schneider10} D. P. Schneider, G. T. Richards, P. B. Hall et al., 2010, \textit{AJ}, \textbf{139}, 2360.
\bibitem{paris15} I. P\^{a}ris et al., 2015, in preparation.
\bibitem{arnaud96} K. A. Arnaud, 1996, \textit{ASP Conf. Series}, \textbf{101}, 17.
\bibitem{antonucci93} R. Antonucci, 1993, \textit{Ann.Rev.Astron.Astrophys.}, \textbf{31}, 473.
\end{thebibliography}
\end{document}